\documentclass[12pt]{iopart}

\usepackage{epsfig}
\usepackage{graphicx}

\begin{document}

\title{Why SETI Will Fail}\footnote{This article originally appeared in the September/October 2002 issue of Mercury magazine (published by the Astronomical Society of the Pacific).}

\author{B. Zuckerman$^1$}

\address{$^1$Department of Physics and Astronomy, University of California, Los Angeles, CA 90095, USA}
\eads{\mailto{ben@astro.ucla.edu}}
\begin{abstract}
The union of space telescopes and interstellar spaceships guarantees that if extraterrestrial civilizations were common, someone would have come here long ago.  
\end{abstract}
\maketitle

\section{Introduction}
Where do humans stand on the scale of cosmic intelligence? For
most people, this question ranks
at or very near the top of the list
of "scientific things I would like to know." Lacking hard evidence to constrain
the imagination, optimists conclude that technological civilizations far in advance of our own are common in our Milky Way Galaxy, whereas pessimists argue that we Earthlings probably have the most advanced technology around. Consequently, this topic has been debated endlessly and in numerous venues.

Unfortunately, significant new information or ideas that can point us in the right direction come along infrequently. But recently I have realized that important connections exist between space astronomy and space travel that have never been discussed in the scientific or popular literature. These connections clearly favor the more pessimistic scenario mentioned above.

Serious radio searches for extraterrestrial intelligence (SETI) have been conducted during the past few decades. Brilliant scientists have been associated with SETI, starting with pioneers like Frank Drake and the late Carl Sagan and then continuing with Paul Horowitz, Jill Tarter, and the late Barney Oliver. Even with all their accumulated talent, these investigators have failed to consider the full implications for SETI of all advanced civilizations possessing space telescopes capable of discovering nearby living worlds. A very likely consequence of such discoveries will be interstellar travel to investigate the nature of alien life forms. The fact that, evidently, no technological creatures bave come to investigate Earth during the past several billions of years is strong evidence that few such creatures exist in our galaxy

\section{Identifying Living Worlds,}
Detection of extrasolar Jupiter-like planets and brown dwarfs is all the rage these days. But to find out more about where we rank on the cosmic intelligence scale, we need to learn about the preponderance and properties of extrasolar Earth-like planets. To do this, we need to launch moderate-sized telescopes into space. These space telescopes will probably be interferometers, although scientists and engineers are currently batting around a variety of designs.

According to the timetable envisioned for NASA's Terrestrial Planet Finder (TPF) mission, in the next 20 years we should witness the deployment of space telescopes capable of spotting Earth-like planets orbiting nearby stars (see Figure 1 and Appendix A: "Terrestrial Planet Finder''). As currently conceived, these telescopes will be able to measure mid-infrared spectra of planetary atmospheres and detect molecules such as water vapor, carbon dioxide, methane, and free oxygen in the form of ozone. In other words, TPF will be capable of identifying life-bearing planets within about 30 light-years of Earth. When technology improves during the coming centuries, the range of such telescopes will no doubt be extended out to 100 or more light-years.

Less than 50 years will separate the beginning of serious SETI efforts and the construction of space telescopes able to detect and study nearby Earths. We therefore live in a unique moment of human history in which we possess powerful ground-based radio telescopes but no TPF. We are an infant technological civilization in this 50 year period, which is an infinitesimal time interval measured in any cosmic context. If we can build TPF so soon after becoming technological and beginning SETI, then other technological civilizations should have little trouble building their own versions of TPF. And SETI pessimists and optimists alike agree that if technological civilizations are numerous in the Milky Way, then a typical civilization must live for a very long time, on the order of millions of years.

Therefore, SETI endeavors should assume that any technological civilization within a few hundred light-years has had space telescopes capable of detecting and studying Earth for quite some time. If the typical technological civilization is 1 million years old, then such a civilization, if it lies within a few hundred light-years, has been studying us with its space telescopes for the past million years. This article will consider some implications of this basic idea.

At present, various radio and optical SETI programs, inclnding the sensitive, ambitious, and relatively expensive Project Phoenix of the SETl Institute, are targeting nearby Sun-like stars. Project Phoenix, led by Jill Tarter, is searching 1,000 nearby stars using telescopes at a pair of widely separated radio observatories to help discriminate against human-made interference. The Phoenix search primarily uses the 305-meter Arecibo telescope in Puerto Rico and the 76-meter Jodrell Bank telescope in Britain.

Rather than hunt blindly for signals coming from nearby stars, the Phoenix team could use its telescope time more economically if it skipped over such stars. Instead, Project Phoenix should target more distant stars that will remain out of range of TPF and its immediate successors $-$ for example stars in the plane of the Milky Way or in the Andromeda Galaxy. Then, a few decades from now, radio and optical SETI programs could intensively focus upon promising planetary systems identified by TPF.

\section{Is E.T. Passive?}

Even if TPF discovers a favorably arranged planetary system or, better yet, a living planet whose atmospheric composition resembles that of Earth, the chance that the plauet hosts a technological civilization is miniscule. I'll explain why.

Our nearest stellar neighbors have been and will remaiu our nearest neighbors for a million years or longer. Because the Sun is traveling at about 10 kilometers per second with respect to most nearby stars, it will take millions of years for Earth to move 100 light-years with respect to these stars.

Suppose that a million years ago Earth entered the sphere of detectability of the space telescopes of an advanced civilization on a planet now within 100 lightyears of Earth. Using their own version of TPF, the inhabitants would have discovered Earth and the fact that it is a life-bearing planet..They would then have pointed some large telescopes at Earth and tried SETI. But the search would have been fruitless because we technological humans were still a million years in the future.

After decades or centuries of unsuccessful SETI, these extraterrestrials would have a decision to make. They might have decided to sit passively as our planetary system drifted by theirs. Such passivity is implicit in a SETI search scheme outlined by Andrew Howard and Paul Horowitz in which, ironically, TPF would be used to detect "deliberate laser transmissions from a technologically advanced civilization" within about 50 light-years of Earth. But these authors fail to consider that the advauced civilization surely would have discovered our living Earth long, long ago, and Howard and Horowitz certainly failed to consider the likely consequences of such a discovery.

Advanced extraterrestrials would have a more attractive and much more plausible alternative to long-term passivity: They could send an interstellar spaceship that contained themselves, or robots, or both, to explore onr living Earth. Everything we know about human nature and history indicates that intelligent creatures will follow this path. Exploration of our solar system began with telescopic observations from Earth. But as soon as we developed the capability, we launched spaceships to explore planets and moons up close because observing from afar is limited and, ultimately, uusatisfying. Without going there, how will we ever find out whether there is or ever was life on Mars, Europa, or Titan?

Some SED proponents, notably Frank Drake and Barney Oliver, have disparaged interstellar spaceships as being slow and expensive compared with radio waves. But intelligent beings aren't going to sit around their home planet for millions of years beaming radio waves into the galaxy. They are going to venture out and explore the universe around them. After all, redwood trees, dinosaurs, and whales do not transmit radio waves. So, if we, or technological extraterrestrials, want to discover and study alien life forms, we must physically travel between the stars in spaceships. How else can we ever know if all life is constructed from proteins and nucleic acids, or is carbon based, or uses liquid water for a solvent? If we do not undertake such voyages, then we will forever forfeit the possibility of answering such profound questions for all living worlds that lack a technological civilization, as Earth did for billions of years.

\section{Interstellar Travel}

Scientists and nonscientists alike are curious about life in the universe. For example, biologist Penelope Boston, a member of the Mars Society's Board of Directors, stated a few years ago on the Discovery Channel's program Destination Mars: "I am a biologist; I have a burning need to know about life in the universe." In 1998, I participated in a NASA sponsored meeting at Caltech on "Robotic Interstellar Exploration in the 21st Century." The engineers and physical scientists in attendance agreed that if humans decide to fund an interstellar mission, which will cost more than all previous space missions, the motivation will be the prospect of investigating a living world at the end of the voyage. Note that a mere 40 years after Sputnik, NASA was already considering the design, cost, and scientific yield of an interstellar mission (see Appendix B: "Interstellar Travel Is Inexpensive").

For interstellar travel to become practical, voyages should take a few hnndred years or less, with perhaps 1,000 years as an upper limit. Scientists and engineers have proposed a variety of propulsion schemes (such as nuclear bombs, pellet streams, and lasers) that could accelerate a spaceship to a few percent of the speed of light. If living worlds are so uncommon in the solar vicinity that the nearest one is farther away than 100 light-years, then exploratory voyages over snch long distances will take such a long time that civilizations will probably send robots and frozen embryos in preference to creatures of flesh and blood (or whatever they're made of). Should living worlds be rare, widely separated, and thus difficult to reach with spaceships, then communicative technological civilizations will be rarer still and SETI searches of nearby stars will fail.

But even if living worlds are not rare, SETI searches of stars within a tew hnndred light-years are doomed to fail because an advanced civilization on any nearby planet would have long ago employed space telescopes to identify Earth as a living planet and would have come to our solar system to investigate Earth. And once here, why leave? Just as the Polynesians who discovered Hawai'i after a long and dangerous voyage across the Pacific did not turn around and return to their point of origin, so representatives of a spacefaring civilizatiou will remain in the planetary systems they choose to explore. After all, it's a long way home.

Thus, the only SETI strategy that makes any sense is to search for signals from distant civilizations, where "distant" is defined as far enough away that Earth has never been discovered as an interesting place by the space telescopes of a putative civilization. Signals received from such distant beings are unlikely to have been generated for our henefit, so detection of extraterrestrial intelligence will likely require luck, and round-trip electromagnetic communications will be slow.

In summary, three simple and plausible postulates have major implications for SETI. First, soon after the development of technology, all civilizations will build space telescopes capable of measuring the atmospheric compositions of Earth-like worlds at distances of hundreds of light-years. Second, intelligent life is curious about other life forms, whether or not that other life is technologicaL And third, once having used space telescopes to discover a nearby living planet, most if not all technological civilizations will be sufficiently curious to construct interstellar spaceships to visit that planet. If these postulates are true, the absence of intelligent aliens in our solar system is strong evidence that they do not exist anywhere in our region of the Milky Way and SETI searches of nearby stars are destined to fail.

\section{Pessimists Are Optimists and Vice Versa} 

Regarding the frequency of technological civilizations in our galaxy, the marriage of TPF and interstellar travel may be extended to distant times and places. Oxygen built up in Earth's atmosphere about 2 billion years ago. Following that period, Earth could have been identified as a living world from afar. Any technological civilization that came within a few hundred light-years of Earth during the past 2 billion years would have had to choose between passively floating by (for a million years) and never learning about terrestrial life, or actively sending a spaceship to our solar system.

Even if such an expedition took 1,000 years, this still would have been a very quick trip in comparison to a billion year wait for humans to show up with radio transmitters. During the past 2 billion years, millions of Sun-like stars have passed within a few hundred light-years of Earth, yet there is no evidence that technological extraterrestrials have ever visited our solar system (see Appendix C: "But What About the Zoo Hypothesis?"). This suggests that very few, if any, technological civilizations existed around these millions of stars. Perhaps the origin of life on Earth was a once-in-a-galaxy fluke, or perhaps life almost never evolves to high intelligence, or perhaps civilizations destroy themselves soon after they develop technology. Although technological life must be exceedingly rare, we currently don't understand astronomy or especially biology well enough to know why.

The above picture, which can be extrapolated to the Galaxy as a whole, is completely different from the one painted by William Newman and Carl Sagan in their ambitious 1981 paper "Galactic Civilizations: Population Dynamics and Interstellar Diffusion." There they model the physical dispersal of technological civilizations as a very slow diffusion process for the following reason. Consider a spacefaring technological civilization arising some 200 light-years from Earth. Then, according to Newman and Sagan, "such a civilization will have been intensively occupied in the colonization of more than 200,000 planetary systems before reaching Earth, some 200 light-years away."

Sagan presented a similar argument in his 1980 book Cosmos, where he remarked on page 308, "that an advanced interstellar spacefaring civilization would have no reason to think there was something special about the Earth unless it had been here already ... From their point of view, all nearby star systems are more or less equally attractive for exploration or colonization." These statements ignore the power of TPF. Just as ground-based telescopes pointed the way to intensive spacecraft exploration of the most interesting planets and moons in our solar system, so would a TPF show a technological civilization just which of 200,000 nearby star systems are worthy of direct exploration.

In conclusion, the relevance of TPF to the question
posed at the beginning forcefully drives home the limited predictive powers of scientists and engineers when forecasting our technological future. Like everyone else, Newman and Sagan failed to anticipate the possibility of TPF and its importance in the SETI debate, even though they wrote their paper just a mere decade before TPF became big news in the space astronomy community. This illustrates how foolishly presumptuous it is when SETI proponents exclaim that interstellar travel is impossible or so difficult that it will never happen. "Never" is a very long time. Already, four spaceships (Pioneers 10 and 11, Voyagers 1. and 2) are exiting the solar system less than a century after the airplane was invented.

The ultimate Luddites are those who deny that human destiny is to venture into space, to become the Little Green Men and Women. Rather than being optimists, SETI proponents who deny this future to human beings are the ultimate technological pessimists. While SETl skeptics may envision humans as possessing the most advanced brains in the Milky Way, nonetheless, it is we pessimists who are the true technological optimists.




\appendix

\section{Terrestrial Planet Finder (by James Larkin of UCLA)}


The Terrestrial Planet Finder (TPF) is an ambitious
proposed NASA space mission to detect and
characterize Earth-like planets orbiting nearby stars.
Currently scheduled for launch in 2014, TPF will search
for planetary systems out to at least 30 light-years. The
exact numbers of targets and distances will depend on
TPFs final architecture. It will be required to produce a
low resolution spectrum with sufficient sensitivity to
identify the basic constituents of a planet's atmosphere.
In particular, TPF can look for spectroscopic features of
water, carbon dioxide, methane, and ozone, which
would be indicative of life.

TPF's enormous challenge can be appreciated when
one realizes that Earth is 10 billion times fainter than the
Sun at optical wavelengths and would be separated by
only 0.1 arcsecond if observed from a distance of 30
light-years (0.1 arcsecond corresponds to half a millimeter
as seen from 1 kilometer). To discern such a faint
object in such close proximity to a star's brilliant glare,
NASA has developed a road map that includes several
precursor missions and technology demonstrations.

All concepts for TPF are space-based, but the final
design is far from decided. Even the operating wavelength
range is still under dehate. The mid-infrared
offers an important advantage because Earth radiates
most of its own light near 10 microns and is only 1 million
times fainter thau the Sun at this wavelength. But
spatial resolution is ultimately set by diffraction effects
that become worse linearly with longer wavelengths. So
although Earth is brighter in the infrared compared to
the optical, an infrared telescope would need to be much
larger to separate Earth from the Sun.

\begin{figure}
\includegraphics[width=150mm]{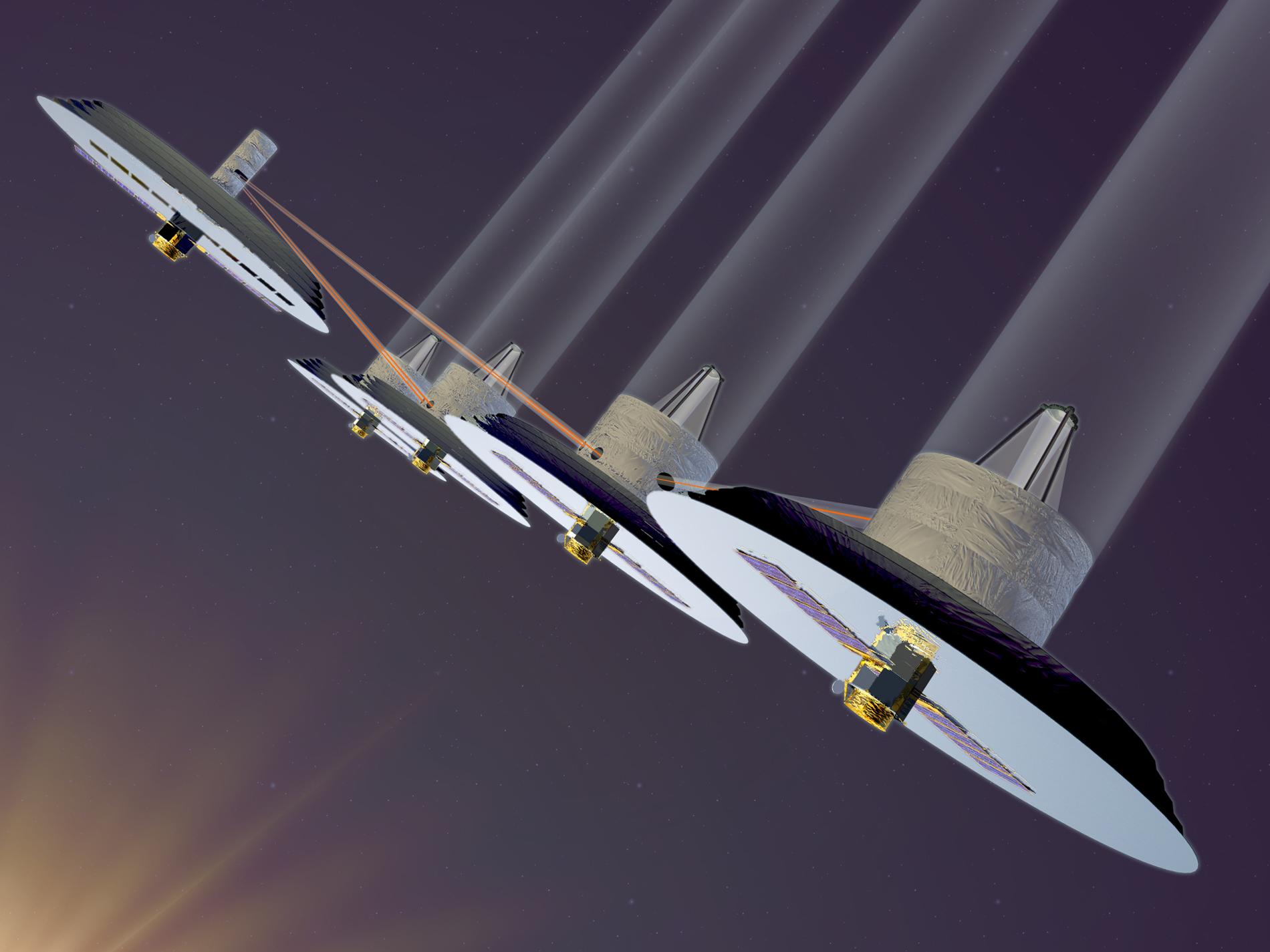}

\caption{\label{figure 1} Artists conception of a space interferometer designed for detection and characterization of living worlds}
\end{figure}

For most of TPF's development, NASA has been considering a mid-infrared interferometer. An interferometer comhines the light of several relatively small telescopes to achieve the angular resolution (with some caveats) of a much larger telescope.

A planet-detecting interferometer works in a nulling mode, which is somewhat different from the operating mode of traditional interferometers. Instead of combining the light from the telescopes in phase, which would increase the measured brightness of the central source, the light is combined out of phase, which suppresses light from a source directly in front of the telescope. But the nulling doesn't work for sources slightly off-center, so a planet would still produce a signal.

The original concept contained two telescopes mounted on opposite ends of a large boom. For planet detection, however, the interferometer must cancel out the star to about 1 part in a million, which requires not only a very good central null but also a null that covers the entire disk of the star. Michael Shao of JPL and Roger Angel of the University of Arizona independently showed in 1990 that the use of four or more telescopes will produce a wider null with greatly improved performance (see Figure A1).

Great strides have already been taken in developing optical and infrared interferometers both on the ground and in space. NASA has awarded contracts to several aerospace companies to develop TPF concepts. Potentially viable alternatives to an interferometer have also emerged. TPF could be a large optical space telescope with a coronagraph to block a star's light. It could also consist of an optical space telescope with a separate free flying occulting mask. Or it could be a specially apodized telescope. The European Space Agency is developing a competitive mission concept called DARWIN. We expect an agreement linking TPF and DARWIN, possibly with Japanese participation as well, in the next several years.

NASA's plans do not end with TPF. Several follow-up missions, most notably Life Finder and Planet Imager, promise ever greater insights into nearby planetary systems. These ambitious programs not only offer crude detection of a few tracer atmospheric molecules; they may also produce rough images of planet surfaces showing at least the presencc of continent- and ocean-sized regions, and they may be able to identify unambiguously the chemistry of a living biosphere. As mentioncd in the main article, all of these missions may be possible less than 100 years after the first large radio telescopes. Similarly, a relatively nearby civilization with its own Planet Imager could have seen Earth's continent-ocean dichotomy. It could have seen annual seasonal variations, the coming and going of ice ages, and long-term changes in vegetation patterns, both natmal and human induced.

\section{Interstellar Travel Is Inexpensive}

SETI proponents often describe the apparently high cost of human space travel to explain how the Milky Way Galaxy could contain numerous technological civilizations while none are present in our solar system. But this perception of "How expensive!" is true only because human "civilization" is seriously out of whack. Should we manage to evolve to a sustainahle society before we totally wreck the biosphere, then human space travel will be inexpensive, as the following comparisons clearly demonstrate.

Consider first a human mission to Mars, a wonderful project that would enlighten the spirit and knowledge of all humanity. According to leaders of the Mars Society, such a mission would cost around 30 billion U.S. dollars. This may sound expensive, but when one realizes that in a typical year the United States spends 10 times this amount on its military budget, and such military expenditures go on year after year after year, by comparison 30 billion is not much money at all. Unfortunately, a human mission to Mars is very unlikely in the foreseeable future. This huge imbalance in spending between the military, which is wasteful at best and destructive at worst, and something so positive as exploration of Mars, is a sad commentary on our couutry and the world.

Of course, sending a spaceship containing humans to another star will cost a lot more than sending a few astronauts to Mars. University of Illinois engineer Cliff Singer, in his excellent chapter "Settlements in Space, and Interstellar Travel" in Extraterrestrials: Where Are They? (a book I co-edited with Michael Hart) estimates the cost of an interstellar spacecraft propelled by a stream of very high-velocity pellets. The estimated price is 1 million person-centuries (l million people each working for a century), or roughly 10 trillion dollars. Expensive? Not really, when one considers that this is about the cost of one decade of the worldwide arms race. Is human "civilization" insane or what?

Various other forms of advanced propulsion systems are discussed in chapters by Freeman Dyson and Ian Crawford in Extraterrestrials: Where Are They?

\section{But What About the Zoo Hypotllesis?}

A well-worn idea in both the scientific and science fiction literature is that "They" are indeed out there, but They are purposely remaining hidden from our view while They keep us under surveillance. In essence, we live in their zoo or wilderness preserve. This "Zoo Hypothesis" implicity assumes that we humans play center stage, otherwise what would They be hiding from?

But as discussed in the main article, if They are indeed out there, the first indication that Earth is an interesting place would have been revealed to them long before we arrived on the scene. Just as the creatures in Sydney, Australia's zoo cannot he observed from New York City, so terrestrial life cannot be observed from even the closest stars. Extraterrestrials must come to our plauetary system before they can even begin to think about zoos. Once in the solar system, they will have no reason not to set up shop in the most obvious places, because there is no need to hide from dinosaurs, lizards, trees, insects, flowers, and fish. By the time technological humans appear, the alien space travelers would be sprawling about our solar system and very much in plain sight.

The above argument, which shows that the Zoo Hypothesis should not be taken seriously, is equally applicable to UFOs. Alien creatures would have arrived in our solar system long before technological Homo sapiens, so it makes no sense for them to be furtively darting about playing hide-and-seek with humanity.

Then there are the related questions: Would aliens send robots or would they come themselves? In the distant or not-so-distant future, will robots constructed by humans supplant humans as the dominant intelligence on Earth?

Answers to these questions are not clear cut, but I think that the creatures themselves will come for the following reason. Whether one views a 3D IMAX film on life aboard the International Space Station versus being on the station, or an IMAX film showing someone climbing Mt. Everest versus actually climbing Everest oneself, the films can never match the adventure of in-person exploration and discovery. When I was a youngster growing up in the Northeast, I was excited by picture books of the wonderful desert canyons of the Southwest. As remarkable as these pictures were, they could not begin to hold a candle to the thrills I felt as an adult while exploring in detail canyons in southern Utah and northern Arizona. No matter how smart our robots get, humans are not going to want to miss an opportunity, to visit other living worlds.



\end{document}